\documentclass[aps,preprint,prl,amsmath,superscriptaddress,floatfix,showpacs,showkeys]{revtex4-1}
\usepackage{graphicx}
\usepackage{dcolumn}
\usepackage{bm}
\usepackage{color}
\usepackage{amsmath}
\usepackage[normalem]{ulem}

\begin{document}
\title{Relationship between Rheology and Structure of Interpenetrating, Deforming and Compressing Microgels}

\author{Gaurasundar M. Conley}
\affiliation{%
 Department of Physics, University of Fribourg, Chemin du MusÃÅ¡e 3, 1700 Fribourg, Switzerland.\\}%
\author{James L. Harden}
\affiliation{%
 Department of Physics, University of Ottawa, Ottawa, Ontario K1N 6N5, Canada\\}%
\author{Frank Scheffold}%
 \email{Frank.Scheffold@unifr.ch}
\affiliation{%
 Department of Physics, University of Fribourg, 1700 Fribourg, Switzerland.\\}%

\date{\today}
\begin{abstract}
Thermosensitive microgels are widely studied hybrid systems combining properties of polymers and colloidal particles in a unique way. Due to their complex morphology their interactions and packing, and consequentially the viscoelastcity of suspensions made from microgels, are still not fully understood, in particular under dense packing conditions. Here we study the frequency-dependent linear viscoelastic properties of dense microgel suspensions in conjunction with an analysis of the local particle structure and morphology based on superresolution microscopy. By identifying the dominating mechanisms that control the elastic and  dissipative response, we propose a unified framework that can explain the rheology of these widely studied soft particle assemblies from the onset of elasticity deep into the overpacked regime. Our results clarify the transition and coupling between the regime dominated by fuzzy shell interactions and the one controlled by the densely cross-linked core. 
\end{abstract}

\maketitle

Soft polymer microgels are fascinating systems whose peculiar properties have resulted in highly diversified applications, spanning from purely academic to the industrial domain \cite{fernandez2011microgel,nayak2005soft}. 
Microgels as model soft spheres have been instrumental in shedding light on fundamental problems relating to phase transitions \cite{paloli2013fluid,zhang2009thermal,mohanty2015multiple} and microgel additives as rheological modifiers are ubiquous in consumer and personal care products as well as other industries \cite{vlassopoulos2014tunable}. The complex nanoscale architecture and softness sets them apart from more conventional solid particles, emulsion droplets or foam bubbles, with profound consequences for the mechanical properties of dense microgel suspensions which reveal rich and complex features in their concentration dependence
\cite{yunker2014physics,vlassopoulos2014tunable}. 
\newline \indent The rheology of hard spherical particles in suspensions is controlled by the volume fraction $\zeta$ of the dispersed phase as the sole parameter determining a suspension's phase behavior. 
In a disordered suspension of hard spheres the maximal volume fraction is reached at random close packing or jamming, $\zeta_\text{rcp}\simeq \zeta_{J}  \simeq$ 0.64 \cite{o2003jamming,hunter2012physics,zhang2009thermal}. Emulsions, bubbles and other soft building blocks, on the other hand, can deform, allowing for ${\zeta _{{\rm{J}}}} \le \zeta  < 1$. In this range the particles form flat facets at contact points which in turn store elastic energy \cite{lacasse1996model,katgert2013jamming,braibanti2017liquid}, resulting in familiar soft pastes such as mayonnaise or shaving foam. 
Polymer microgels however are different. They are highly swollen in good solvent conditions, 
and as a consequence, microgels are compressible in addition to being deformable and therefore highly overpacked states can be reached 
\cite{rovigatti2017internal,senff1999temperature,pelton2000temperature,fernandez2011microgel}. Moreover, microgels prepared following standard protocols have a fuzzy shell decorating their compressible cores \cite{stieger2004small,reufer2009temperature}, allowing for shell compression and interpenetration \cite{mohanty2017interpenetration}. Much work has been devoted to the characterization of the rheology of dense microgel suspensions (or pastes) \cite{senff1999temperature,senff2000influence,stieger2004thermoresponsive,sessoms2009multiple,lietor2011bulk,menut2012does} and common features for microgels of different sizes and softness have been established. The elastic modulus grows rapidly after the liquid-solid transition and then much more slowly at higher concentrations. Although models have been developed to account for this behavior, there still exists no widely accepted framework that encompasses the entire range of packing densities. One of the main reasons for this unsatisfactory situation is that, in the past, little or no in-situ information on the single-particle nanoscale level has been available. Recently, however, significant progress has been reported in studies revealing single particle properties in dense suspensions based on zero average contrast small angle neutron scattering \cite{mohanty2017interpenetration} and microscopy \cite{conley2016superresolution,conley2017jamming,de2017deswelling,bergmann2018super}. 
\newline \indent In this work, we propose a framework to explain the frequency-dependent linear viscoelasticity of dense microgel suspensions from weak packings to strongly overpacked states by combining the results from oscillatory shear measurements and nanoscale imaging. To this end we characterize the macroscopic rheological properties at a constant temperature and take advantage of the advent of microscopic structural information about individual microgels and microgel pairs resolved via superresolution microscopy.
Our aim is to describe and connect the mechanisms that dominate the elastic and lossy behavior across the different concentration regimes. 
\section{Results}
\subsection{dSTORM superresolution microscopy}  
\begin{figure}[ht]
\centering
  \includegraphics[trim={.3cm 0cm .5cm 0.2cm},clip, width=1\columnwidth]{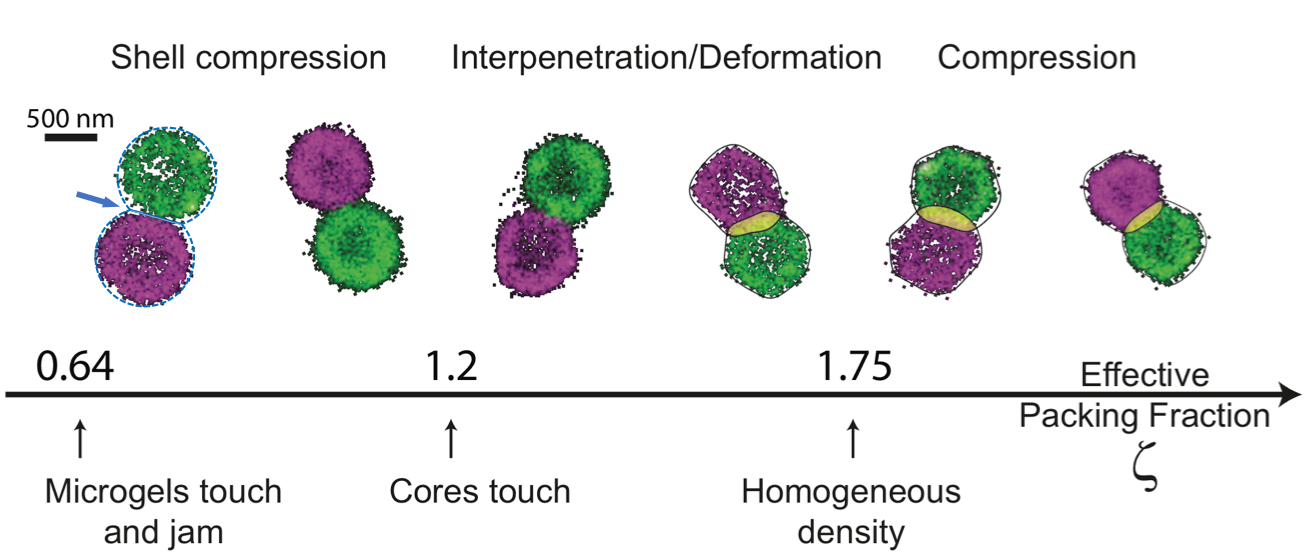}
  \caption{Two-color dSTORM superresolution microscopy of dye labeled tracer particles seeded in densely packed microgel suspensions \cite{conley2017jamming}. The effective packing fraction $\zeta$ increases from left to right ($\zeta=0.86,1.01,1.26,1.50,1.89,2.13$).  
Left: The dashed circles with radius $R_\text{tot}=470$nm  at $T = 22 ^{\circ}$C visualize the total microgel size including the barely visible low density corona. The arrow points to the contact area where the brush-like corona is partially compressed. The straight line indicates the cross section of the contact area. Right: The  solid lines show the contour of the microgels for higher packing densities where the corona has already been fully compressed onto the core and microgels interpenetrate \cite{conley2017jamming}. The overlap area $\Delta F$ is highlighted in yellow.}
  \label{fgr:stormrecap}
\end{figure}

We study pNIPAM microgels prepared by free radical precipitation polymerization as described in \cite{senff1999temperature,conley2016superresolution}.   The experimentally accessible mass concentrations $c$ of our suspensions, in wt/wt \%, can be converted into effective packing fractions $\zeta=k\times c$ via the voluminosity $k=0.08$, as shown in \cite{conley2017jamming}.  We have verified, using small angle light scattering that, that on the time scale of the experiment, the samples do not crystallize \cite{conley2017jamming}. The structure and morphology of standard, submicron sized Poly(N-isopropylacrylamide) (pNIPAM) microgels can be resolved via single and dual color superresolution microscopy, from marginally jammed to deeply overpacked states, as depicted in Figure \ref{fgr:stormrecap} \cite{conley2016superresolution,conley2017jamming}. We have identified three consecutive stages of packing as descried earlier \cite{conley2017jamming}. In the first stage, just above solidification ($\zeta\gtrsim 0.64$), we observe some mild compression of the microgel's fuzzy corona. 
In the following stage, once the dense cores come into contact ($\zeta\gtrsim 1.1$), interpenetration becomes noticable and the microgels start to significantly deform, without changing in size. Interpenetration gradually increases as the contacting facets expand. Finally, once interpenetration and deformations have saturated and the volume is homogeneously filled by the polymer gel ($\zeta\gtrsim 1.75$), isotropic compression remains the only mechanism that allows further densification. 

\subsection{Oscillatory shear experiments} 
\begin{figure}[ht]
\centering
\includegraphics[width=1\columnwidth]{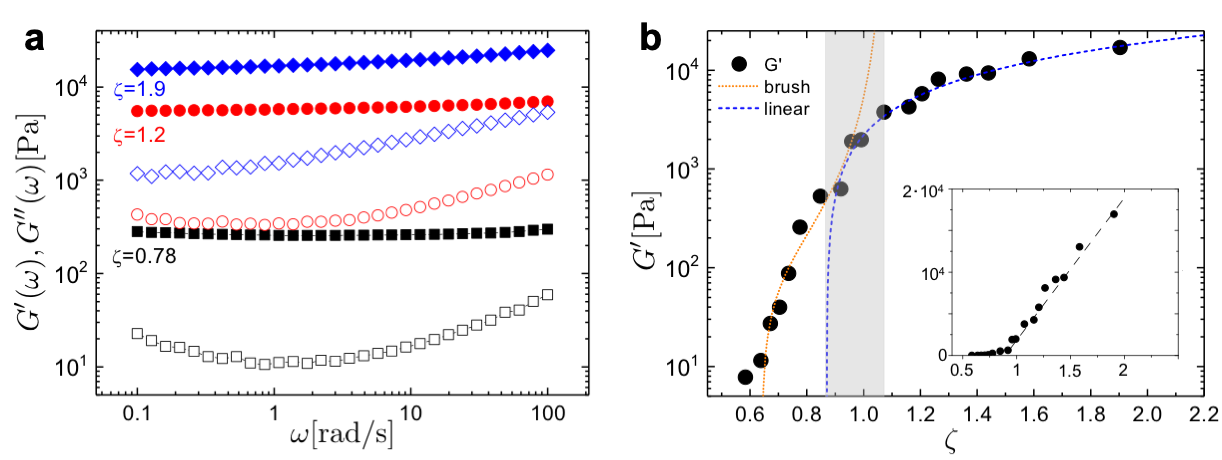}
  \caption{(a) $G^\prime(\omega)$ (full symbols) and $G^{\prime\prime}(\omega)$ (open symbols) as a function of frequency $\omega$ for different packing fractions, ranging from marginally jammed to deeply overpacked. (b) $G^\prime(\omega=1.2 \text{rad/s})$ as a function of packing fraction $\zeta$ fitted with the Brush model (dotted line) at lower $\zeta$ and with a linear function $G^\prime = 1.7 \text{kPa} \times (\zeta-\zeta_{c})$, dashed line for $\zeta\ge \zeta_{c}=0.87$. Inset: Linear representation of the same data.}
  \label{fgr:g12}
\end{figure}

We perform oscillatory shear measurements in the linear regime (strain $\gamma = 0.1\%$) at a fixed temperature of $T = 22 ^{\circ}$C covering a wide range of $\zeta$, from the onset of jamming  $\zeta \gtrsim \zeta_J=0.64$ to deeply overpacked, and determine the elastic and loss moduli as a function of frequency, $G^\prime(\omega)$ and $G^{\prime\prime}(\omega)$. Selected examples of frequency dependent measurements of $G^\prime$ and $G^{\prime\prime}$ are shown in Figure \ref{fgr:g12}(a) covering the $\zeta$ range from marginally jammed to deeply overpacked. 
In all cases we find $G^\prime$ being nearly frequency independent and greater than $G^{\prime\prime}$, indicating solid like behavior.
Losses, however, are relatively high and $G^{\prime\prime}(\omega)$ shows a minimum around $\omega \sim 1$ rad/s, typical of emulsions and foams, in addition to microgels \cite{lacasse1996model,gopal2003relaxing,lietor2011bulk}. 
With increasing concentration the minimum becomes progressively less pronounced and finally, with $\zeta = 1.9$, has all but disappeared. 
To characterize the $\zeta - $dependent elasticity of our microgel suspensions we take the value $G^\prime(\omega)$ at a fixed frequency of $\omega = 1.2 $ rad/s, 
Figure \ref{fgr:g12} (b). Below the jamming packing fraction $\zeta_J$ we measure a weak elastic modulus $\sim k_{B}T/R^3$ that we can tentatively ascribe to the entropic glass regime but that is barely resolved in our experiments \cite{ikeda2013disentangling,scheffold2013linear,pellet2016glass,braibanti2017liquid}. Starting from $\zeta\sim \zeta_J$, when microgel coronas are in direct contact, we find a steep increase of $G^\prime$. Increasing $\zeta$ by a factor two results in a three order of magnitude increase of $G^\prime$. This trend however does not continue over the entire range, instead we observe a slow crossover into a different regime where the slope is reduced considerably.  
\section{Discussion}
\subsection{Elasticity and Storage modulus $G^{\prime}(\omega$)}
In the past the rapid increase of elasticity after jamming has been described \emph{ad-hoc} in terms of a soft interaction potential of the form $\psi \sim r^{-n}$ \cite{senff1999temperature,stieger2004thermoresponsive} resulting in a power law $G^\prime \sim \zeta^m$ with $m = 1+n/3$. A more physically descriptive, yet still quantitative approach has been proposed by Scheffold et al. \cite{scheffold2010brushlike}, whereby the microgels are modeled as solid cores, of size $R$, decorated by polymer brushes, of thickness $L_0\simeq R_\text{tot}-R$, which mediate their interactions. The net repulsion between brushes at the microgel periphery, derived from the Alexander - de Gennes scaling model for polymer brushes in good solvent contitions, suffices to describe the onset of solid like behavior. To derive an expression for $G^\prime$, a local spring constant $\kappa$ is defined which which is directly related to the interaction potential between two spheres by $\kappa = \partial ^{2}\psi / \partial  r ^{2}$, and to the elastic modulus as $G^\prime \approx \kappa / \pi R$. The complete expression for $\frac{G^\prime(\zeta,\alpha)}{kT\alpha/s^3}$, modulus a prefactor of order unity, is 
given in \cite{scheffold2010brushlike}
where $s$ is the effective average separation between grafting sites and $\alpha$ is the ratio between core and total radius of the particle.
By setting $\alpha = 0.84$ and adjusting the prefactor $k_\text{B}T\alpha/s^3 =$ 60 Pa for a best fit we obtain the dotted curve shown in Figure \ref{fgr:g12} (b). The value of $\alpha = 0.84$ compares well with the static light scattering result $R/R_\text{tot}\simeq 0.82$ and previous studies on similar systems \cite{scheffold2010brushlike}.  
We find very good agreement with the experimental data in the lower concentration range, $ 0.64 \le \zeta \lesssim 1$, delineating the range where the microgels are predominantly interacting via their brush-like coronas, Figure \ref{fgr:stormrecap}. Instead of the divergence of $G^\prime(\zeta)$, predicted by the brush model we find, at higher concentrations, a slower, linear increase of elasticity as a function of packing fraction.
 This is in agreement with several previous studies on dense microgel packings \cite{pellet2016glass,di2013macro,di2013particulate,menut2012does,romeo2013elasticity} and can be attributed to the finite softness of the microgel core.  
By extrapolation, we can estimate an onset of the linear regime at $\zeta_c = 0.87 \pm 0.01$. This occurs well before the divergence predicted by the brush model at $\zeta \approx 1.08$, resulting in a crossover region where the softness of the core eventually dominates over the stiffness of the highly compressed corona. The value of $\zeta_c$ is consistent with the jamming of the discrete homogeneous particles of size $\sim 0.9 R_\text{tot}$ slightly smaller than the unperturbed radius $R_\text{tot}$ (i.e. when the fuzzy shell has been compressed almost entirely onto the core). 
As $\zeta$ is increased further, 
significant particle deformations can be seen by dSTORM, Figure \ref{fgr:stormrecap}.
The linear scaling of elastic modulus extends to the limit of very high densities where the  systems becomes more and more homogeneous as confirmed by small angle light scattering \cite{conley2017jamming}. 
Interestingly, Calvet et al. found \cite{calvet2004rheological}, for macroscopic pNIPAM gels of similar composition to ours ($\sim$ 5 mol $\%$ BIS), elastic moduli $G^\prime \approx 10^{4} $Pa comparable to our results at $\zeta \geq 1.75$.
\subsection{Friction and loss modulus $G^{\prime \prime}(\omega$)} Next we consider the energy losses in the system where the influence of discrete particulate nature of the suspension at high packing fractions is striking.  Over the entire range we find significant losses, typical for soft glassy materials \cite{chen2010rheology} but in stark contrast to macroscopic gels which are almost entirely elastic in their stress response. Calvet et al. found \cite{calvet2004rheological}, for macroscopic pNIPAM gels $G^{\prime \prime}/G^\prime\sim 10^{-3}$, typically  about two orders of magnitude less than what we observe. Such anomalously large losses are well known for jammed emulsions and they are present despite the fact there are no static friction forces between the emulsion droplets. Liu et al. showed, that the high losses are due to dynamic dissipation in the fluid confined between planes of facets sliding relative to each other \cite{liu1996anomalous}. The random orientation of slip planes leads to a broad range of stress relaxation rates that result in $G^{\prime \prime}\sim A(\eta) \omega^{0.5}+\eta_\infty \omega$ in this regime, where $\eta_\infty$ denotes the background viscosity of the solvent phase. Their amplitudes $A(\zeta)$ increase by about a factor 3-4 over the concentration range explored, $\zeta\simeq 0.6-0.86$, which can be explained by the increased viscosity of the compressed liquid film confined in the shear planes \cite{liu1996anomalous}. 
\begin{figure}[ht]
\centering
\includegraphics[trim={0cm 0.0cm 0cm 0cm},clip,width=1\columnwidth]{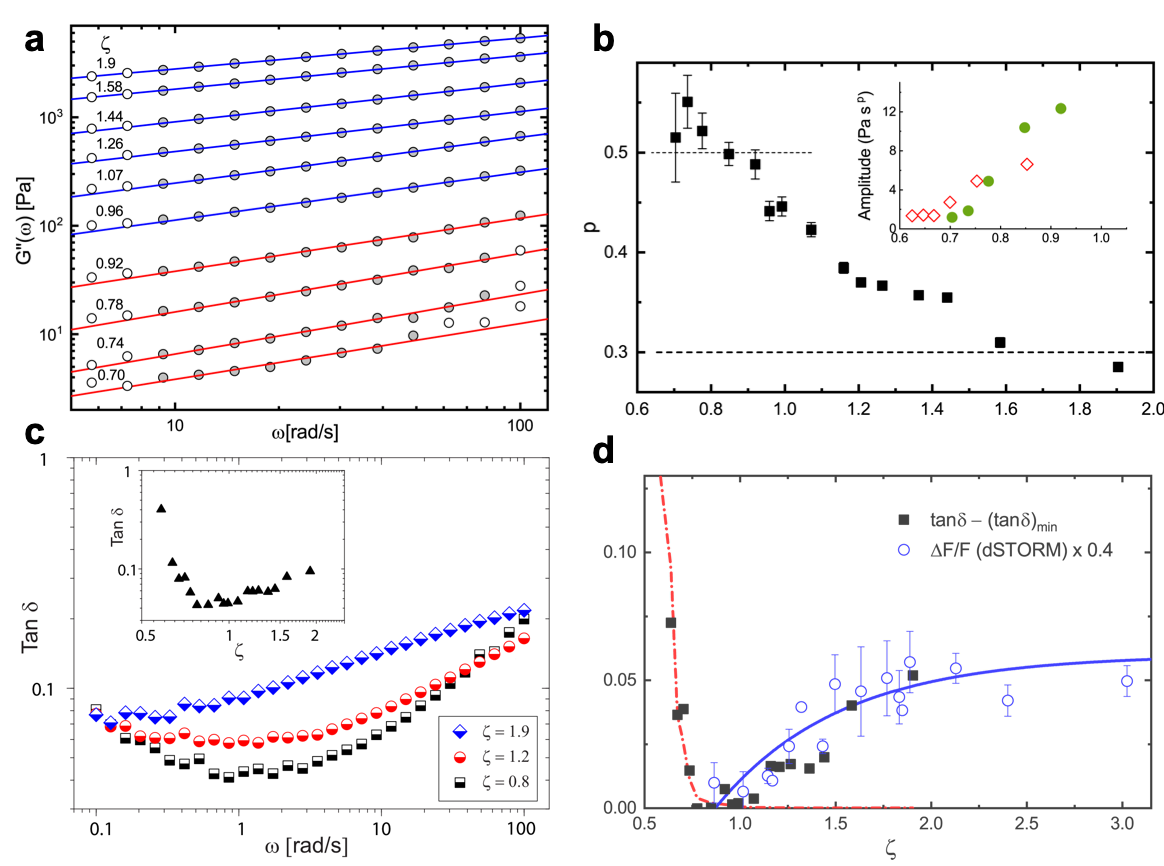}
  \caption{(a) Loss modulus $G^{\prime \prime}$  for frequencies $\omega \ge 10$ rad/s for different packing fractions (symbols). Solid lines show the fit to the data with $G^{\prime \prime}(\omega)=A(\zeta)\omega^p$. Data points included in the fit range are filled dark grey. Lines with $p\simeq 0.5$ are red, $p<0.5$  black. (b) Values of $p$ obtained by a best fit to the data. Inset: full symbols show the concentration dependence of $A(\zeta)$ for microgels.  Open symbols show  data  for emulsion, from ref. \cite{liu1996anomalous} with $p=0.5$. (c) Loss tangent $ \tan \delta$ as a function of frequency $\omega$ for different packing fractions, ranging from marginally jammed to deeply overpacked. Inset: $ \tan \delta$ as a function of packing fraction $\zeta$ for $\omega = 1.2$rad/. (d) Solid squares: $\tan \delta - (\tan \delta)_\text{min}$ with $(\tan \delta)_\text{min}=0.043$. Open circles: rescaled overlap area $0.4 \times \Delta F/F$ of adjacent microgels derived from dSTORM, as shown in Figure \ref{fgr:stormrecap} (full data set reproduced from \cite{conley2017jamming}). Dash-dotted line denotes $1\text{Pa}/G^{\prime}$ and solid line shows $\lambda (1-\exp{[-(\zeta-\zeta_c)/\xi}])$ with $\lambda$ = 0.06 and $\xi = 0.65$. }
  \label{fgr3}
\end{figure}
\newline \indent Based on the emulsion work and by comparison with the dSTORM data we can now verify the accuracy and the range of validity of this scenario for our microgel system. As discussed before, we described the microgel by a cross-linked core covered by a brush-like corona. At $\zeta > \zeta_J$ the brushes are partially compressed and the restoring forces lead to the finite macroscopic shear modulus $G^\prime$, as shown in Fig \ref{fgr:g12} (b). At the same time it is known that compressed polymer brushes do not show any noticeable friction when sheared slowly against each other, down to compression ratios of $L/L_0\sim 0.1-0.15$ in thickness \cite{klein1994reduction}, suggesting that the model developed for emulsions, should also apply to microgels, over a limited concentration range. As a critical test, 
we fit $G^{\prime \prime}(\omega)=A(\zeta)\omega^p$ with adjustable parameters $A(\zeta)$ and $p$ roughly over a decade in frequency $\omega \in [10,100]$ rad/s. Figure \ref{fgr3}(a) shows these fits in a logarithmic representation for each choice of $\zeta$. We note that the contribution of the background fluid is small enough that it's contribution can be safely neglected over the range of frequencies $\omega \le 100$rad/s considered. Up to $\zeta \simeq \zeta_c$ the data is well described by the $\omega^{0.5}$ scaling predicted for emulsion droplets with no static friction and also the amplitudes $A(\zeta)$ are similar to the one reported in ref. \cite{liu1996anomalous}, inset Figure \ref{fgr3}(b). Quantitative differences in $A(\zeta)$ between emulsions and microgels can be explained by the fact that disjoining pressure between the droplet interfaces and the compressed brushes of the microgel corona are not exactly the same. 
\newline \indent 
Figure \ref{fgr3}(b) shows the dependence of the power-law fit parameters $A$ and $p$ on $\zeta$. Starting at $\zeta \sim 0.9$ deviations form the $\omega^{0.5}$ scaling can be clearly observed. For a solid core the corona would be entirely compressed on the core at $\zeta \to 1.08$, but in our case the core and the corona deformation are coupled and the transition is smeared in the range $\zeta \in [0.87,1.08]$ due to the compressibility of the core.  Thus for $\zeta \ge \zeta_c$ we expect the density and osmotic pressure of the corona and the core to gradually approach each other \cite{romeo2013elasticity}. As a consequence the entropic penalty for the dangling ends of the corona to interdigitate becomes smaller and smaller and losses, expressed by the ratio $G^{\prime \prime}/G^\prime=\tan\delta$, increase over the entire spectrum, Figure \ref{fgr3}(c). \newline In this regime, two-color dSTORM provides key information about this interdigitation process. The open circles in Figure \ref{fgr3}(d) show the overlap area $\Delta F/F$ extracted from a $\sim500$nm thick z-section through the center of adjacent microgel particles \cite{conley2017jamming}. The overlap increases rapidly from $\zeta_c=0.87$ to $\zeta=1.9$ and saturates above. 
We also observe a lowering of the high frequency slope, from $ G^{\prime \prime} \sim \omega^{0.5}$ to $\sim \omega^{0.3}$, dashed lines in Fig. \ref{fgr3}(b),
marking a deviation from the viscous behavior of jammed emulsions where the slope of 0.5 is maintained \cite{liu1996anomalous} but in agreement with previous observations on dense microgel suspensions \cite{menut2012does,sessoms2009multiple}.  Interestingly we find that the anomalously large losses, expressed in terms of $\tan\delta=G^{\prime \prime}/G^\prime$, scale directly with the overlap area $\Delta F/F$ derived from superresolution microscopy, as shown in Fig. \ref{fgr3}(c). In particular, both seem to rise together roughly exponentially toward a plateau value at large $\zeta$. We stress the fact that for $\zeta<\zeta_c$ the situation is entirely different. As long as the corona is not yet fully compressed, brush-brush interfaces are lubricated,  $G^{\prime \prime}$ increases slowly and thus the relative viscous losses drop with the modulus: $\tan\delta\propto 1/G^{\prime }$ as shown by the red dash-dotted line in Figure \ref{fgr3}(d).  
\section{Conclusions} Our investigations reveal that the onset of elasticity 
in the dense microgel suspensions is governed by the overlap and compression of their fuzzy outer shells. 
At higher packing fractions, we visually observe deformation and compression and here the elasticity increases linearly with concentration starting at $\zeta_c$, in agreement with the jamming picture of dense assemblies of homogeneous soft spheres \cite{o2003jamming,paloli2013fluid,seth2011micromechanical,pellet2016glass}. 
Moreover, this linear dependence continues into the deeply overpacked regime, $\zeta>\zeta_c$, where it is in accordance with the behavior found for macroscopic gels.
Interestingly, the observed significant and mounting viscous dissipation demonstrates that the particulate nature of microgel suspensions remains dominant up to the highest packing fractions studied. Despite the close similarities with dense emulsions for $\zeta$ between $0.64$ and $0.9$, we find that the microgel systems enter a distinctly different regime with respect to dissipation at higher packing fractions, not accessible to emulsions, due to the effects of the unconnected, interpenetrating corona chains. While storage moduli $G^\prime (\omega)$ are nearly independent of frequency $\omega$, both for emulsions and microgels, up to the maximum possible packing density, the loss spectra $G^{\prime\prime} (\omega)$ of our compressed microgels vary quite significantly with concentration,  increasing even more rapidly with $\zeta$ than the storage modulus above $\zeta_c$.
Indeed, superresolution microscopy suggests that faceting and weak interpenetration opens up new pathways for dissipation which explains the rising loss modulus in the overpacked regime and again highlights the essential role played by the particulate nature of the microgel suspensions and the microgel structure even at the highest packings.
\section{Methods}
\subsection*{Microgel synthesis} The microgels were synthesized by free radical precipitation polymerization in order to obtain micron sized particles as described in \cite{senff1999temperature,conley2016superresolution}.  This standard protocol, used in the vast majority of studies on microgels, is known to produce inhomogeneous particles with a dense core surrounded by a fuzzy shell or corona. Other, more recent synthesis approaches, use starved feed conditions to produce much more homogeneous particles without dangling ends which are not subject of this study and could be addressed in future work \cite{still2013synthesis}. The co-monomer that binds to the fluorescent dye for dSTORM imaging was added with a delay in order enhance the signal from the boundaries leading to a depletion of signal in the center of the particles, as shown in Figure \ref{fgr:stormrecap}, which is insignificant for our analysis. The reconstructed 2D images originate from a plane of $\sim$ 500 nm thickness adjusted to the center of the particles. This follows from our image reconstruction protocol, where we set a corresponding threshold along the $z-$axis \cite{conley2017jamming}.  To enhance the dye label signal of the outer region of the microgels the co-monomer was added continuously to the reaction solution with an $\approx$ 10 minutes  delay after particles have begun nucleating. The latter can be observed visually (the solution becomes milky). Dense samples were obtained by centrifugation of a dilute stock suspension and the the weight concentration of the stock solution  was determined by drying and weighing.  Our swollen microgels have a total radius of $R_\text{tot}=470$ nm (polydispersity 6\%) and a core radius of $R=380$ nm, determined by static light scattering from a dilute suspension at $T = 22 ^{\circ}$C \cite{conley2017jamming}.

\subsection{Rheology}
Measurements were performed on a commercial rheometer (Anton Paar MCR 502), using a cone-plate geometry (cone radius 25 mm, angle 1.0$^{\circ}$), equipped with a solvent trap to limit evaporation during the measurement.

\subsection*{dSTORM superresolution microscopy}
The microgels were labeled with the fluorophore Alexa Fluor\textsuperscript{\textregistered}647. We added 50 mM Cysteamine (Sigma Aldrich) and adjusted the pH to 8 using HCl \cite{conley2016superresolution}. For two-color dSTORM the Cysteamine concentration was increased to 100 mM. Between 60 and 80,000 frames were recorded at 60 to 100 frames per second \cite{conley2017jamming}. From the localization of single fluorophores we extraced the coordinates used to reconstruct a superresolved image  \cite{ovesny2014thunderstorm}. A detailed description of the dSTORM imaging protocol has been published earlier \cite{conley2016superresolution,conley2017jamming}.

\section{Acknowledgements} This work was supported by the Swiss National Science Foundation through projects 149867 and 169074 and benefitted from support by the National Center of Competence in Research \emph{Bio-Inspired Materials}. F.S. acknowledges financial support by the Adolphe Merkle Foundation through the Fribourg Center for Nanomaterials. J.L.H. acknowledges support from NSERC. We would like to thank Veronique Trappe, Philippe Aebischer, Sofi N\"ojd and Peter Schurtenberger for discussions. 
\section{Authors contributions} FS and GMC conceived the study. GMC did all experiments. All authors contributed to the data analysis and writing of the manuscript.

\end{document}